\documentclass[12pt]{article}
\usepackage{a4}
\usepackage{amssymb}
\usepackage{epsf}
\usepackage{cite}
\def\Bbb{\mathbb}
\def\BZ{\Bbb Z} 
 \def\BP{\Bbb P}
\def\BT{\Bbb T} 

\catcode`@=11 \@addtoreset{equation}{section} \catcode`@=12

\begin{document}
\begin{titlepage}
\begin{flushright}
{\tt hep-th/0104126}\\
{\tt IMSc/2001/04/22} \\
April, 2001
\end{flushright}
\begin{center}
{\Large \bf Boundary Fermions, Coherent Sheaves and D-branes on Calabi-Yau
manifolds} \\[1cm]
Suresh Govindarajan\\
{\em Department of Physics, Indian Institute of Technology, Madras,\\
Chennai 600 036, India\\
{\tt Email: suresh@chaos.iitm.ernet.in}\\[10pt]}
and \\[10pt]
T. Jayaraman\\
{\em The Institute of Mathematical Sciences, \\ Chennai 600 113, India\\
{\tt Email: jayaram@imsc.ernet.in}\\[10pt]}
\end{center}
\vfill
\begin{abstract}
We construct boundary conditions in the gauged linear sigma model for B-type
D-branes on Calabi-Yau manifolds that correspond to coherent sheaves 
given by the cohomology of a monad. This necessarily involves the
introduction of boundary fields, and in particular, boundary fermions.
The large-volume monodromy for these D-brane configurations
is implemented by the introduction of boundary contact terms. We also
discuss the construction of D-branes associated to coherent sheaves that
are the cohomology of complexes of arbitrary length.  
We illustrate the construction using examples, specifically those
associated with the large-volume analogues of the 
Recknagel-Schomerus states with no moduli. Using some of
these examples we also construct D-brane states that arise as bound
states of the above rigid configurations and show how moduli can be
counted in these cases. 
\end{abstract}
\vfill
\end{titlepage}

\section{Introduction}
 
As in the case of closed string compactifications on Calabi-Yau
manifolds, the gauged linear sigma model description 
appears to be the natural starting point 
in the context of
D-branes wrapped on supersymmetric cycles in Calabi-Yau manifolds,
particularly for the study of the
dependence of D-brane physics on the K\"ahler moduli.(For 
a list of references on D-branes on CY manifolds see
\cite{RS,quintic,dgepner,oops}.) In an
earlier paper\cite{lsmone}, we had taken the first steps towards such a
description studying in particular the case of the six-brane wrapped on
the full Calabi-Yau manifold. Subsequently, following on the 
work of
\cite{diacgom} and \cite{dougdiac} we had demonstrated a systematic
procedure using the techniques of helices and mutations whereby we could
construct the large volume analogues of all the $\sum_a l_a=0$
Recknagel-Schomerus states as restriction of 
exceptional sheaves in the ambient variety to the Calabi-Yau
hypersurface\cite{helices}.
While the procedure used the intuition of the GLSM we had
not provided an explicit field-theoretic method of construction of these
D-brane configurations in the GLSM. This paper is devoted to such
explicit constructions. 

The paper in essence extends the techniques developed in the context of
heterotic strings to describe vector bundles on Calabi-Yau manifolds to
the case of world-sheets with boundary. In the heterotic case these
essentially involved using the monad constructions of vector bundles in
the case where all the elements in the complex involved only line
bundles. In the first instance in this paper we extend this technique to
the case of worldsheets with boundary.  

But since this provides a
extremely limited class of configurations (and not even the full set of
Recknagel-Schomerus states) we are led naturally
to constructions of D-brane configurations that are described by
complexes of length greater than two. Such constructions are also
important in the light of the role that the derived category of coherent
sheaves\footnote{For some earlier work where the derived category of coherent
sheaves appeared in the context of string theory, see \cite{cat}.}
is expected to play in understanding issues related to
B-type D-branes especially
in the stringy regime as argued by Douglas in \cite{dougtalk,dougcat}.
The derived category of coherent sheaves is precisely the description of
all sheaves on the CY manifold by complexes that are generically longer
than monads. It would be of some importance to extend the construction
involving complexes of length greater than two to the heterotic string.

The description of the formation of bound states of D-brane
configurations in terms of sequences of coherent sheaves, 
as first explained by Harvey and Moore \cite{bpsalgebra}, is easily  
implemented in the GLSM construction. We provide examples of such
description of bound states in the context of the D-brane configurations  
that we consider as examples in this paper.

The organisation of the paper is as follows: In sec. 2 we remark on the
use of boundary fermions in relation to vector bundles. In sec. 3  we
set up our notation and discuss the decomposition of bulk multiplets
under B-type supersymmetry.  Section 4 describes the mathematical
construction of bundles using complexes with an emphasis on the monad
construction. Section 5 gives an explicit realisation, in the GLSM with
boundary, of bundles that
are described by monads in sec. 4. In sec. 6, we discuss how the large
volume monodromy action on the vector bundles is realised in the GLSM.
In sec. 7, we discuss the GLSM construction for arbitrary complexes.
In sec. 8, we discuss how D-branes corresponding to bound states
may be realised by our methods using specific examples. We conclude
in sec. 9 with a brief discussion. 

The possibility of using boundary fermions in the GLSM to describe D-brane
configurations has been discussed in a talk by S.
Kachru\cite{stringstalk}. In the course of our work a paper
appeared \cite{hori} that uses boundary fermions in relation 
to B-type D-branes but in a very limited context. Lastly, as this
manuscript was being completed for publication another paper \cite{mcg}
appeared, that has some overlap with the work of sec. 7 of our paper. 

\section{Some remarks on boundary fermions}

In an earlier paper\cite{lsmone}
 where we discussed the GLSM with boundary, we did not
introduce boundary fermions but nevertheless obtained some consistent
boundary conditions. By consistent boundary conditions, we mean boundary
conditions in the GLSM which have a proper NLSM limit. As was shown in
that context, this imposes a rather stringent condition on the possible
boundary conditions. In that construction two problems arose in the
context of fairly simple examples:
\begin{enumerate}
\item[(a)] One has to find appropriate boundary conditions for the
$p$-field which is set to zero in the NLSM limit. The boundary
condition proposed in \cite{lsmone}
was to impose $p=0$. This is acceptable in the CY-phase but clearly
is in conflict with the bulk ground state condition in the LG-phase.
\item[(b)] For the case of ``mixed''-boundary conditions, i.e., for
branes such as the $D2-$ and $D4-$ brane, the boundary conditions on the
fields in the vector multiplet were somewhat contrived and unnatural
unlike the case of the D6-brane, i.e., the brane wrapping the full CY
manifold.
\end{enumerate}

In this paper we obtain a simple resolution to these two problems by
introducing boundary fermi supermultiplets. The boundary fermions
impose the condition $P=0$ for case
(a) and $f(\phi)=0$ for case (b) as a ground state condition. In the
process one finds that one can obtain natural boundary conditions on the
fields in the vector multiplet from the NLSM limit.

For the case of
coincident D-branes, the boundary fermions can be interpreted as objects
carrying the Chan-Paton index. More precisely, they are objects carrying
indices associated with the vector bundle of the corresponding D-brane.
In the GLSM, these have non-trivial interactions with the bulk
multiplets. We note that boundary fermions were introduced early on
as carriers of the Chan-Paton index by Marcus and Sagnotti\cite{ms}(see
also \cite{andtsyet}).
More recently, they have been introduced in the study of tachyon
condensation and non-BPS states beginning with the work of
\cite{wittenktheory,Kutasov:2000aq}.

With the introduction of boundary fermions, the worldsheet Witten 
index computation becomes completely analogous to
the original calculations of the index theorem using supersymmetric
quantum mechanics. 
Consider the case, when the worldsheet is topologically a cylinder.
We will be interested in the case, when the two boundaries end on
D-branes corresponding to non-trivial vector bundles. The D-brane
configurations are assumed to preserve B-type supersymmetry. The
Witten index associated with the BRST charge $Q\equiv 
\overline{Q}_+ +\eta \overline{Q}_-$ can be seen to be equal to
$$
\chi(E_1,E_2)\equiv \int_M {\rm ch}(E_1) {\rm ch}(E_2^*) {\rm Td} (M)
\quad,$$
(see \cite{HIV}) where $M$ is the target manifold and $E_1$ and $E_2$
are the vector bundles on the two boundaries. 

It is known that the contributions ${\rm ch}(E_1)$ and ${\rm ch}(E_2^*)$
from the two boundaries can be realised by introducing fermions living
on the boundaries. More precisely, the path-ordered integral
$$
P\left(\exp\left[\int dx^0 \partial_0 \phi^\mu A_\mu^r(\phi) T^r\right]\right)
$$ 
(where $T^r$ are in the fundamental representation) is equivalent to
the path-integral of anti-periodic (complex) fermions in the fundamental
representation of $E$ provided we restrict to {\em one-particle
states}\cite{luis}.
The action is given by 
$$
S_{b.f.}=\int dx^0 [\overline{\pi}_a D_0 \pi_a]\quad,$$
where $D_0 \pi = (\partial_0 + \partial_0 \phi^\mu A_\mu^r(\phi)
T^r)\pi$.
 If one however restricts to $n$-particle states, then the path integral
leads to gauge fields on the $n$-th anti-symmetric power of $E$. On the
other hand, if one allows all states, then one obtains gauge fields on
the {\em spinor bundle} over $E$\footnote{We thank K.~Hori for bringing
this issue to our attention.}.

In what follows in the rest of the paper, we will assume that such a
restriction to one-particle states is in operation always (this can be
trivially done by using a Lagrange multiplier\cite{dhoker})except
where we explicitly consider situations to the contrary.   

\subsection{Boundary fermions and the GLSM}

For supersymmetric D-brane configurations, the boundary preserves
a linear combination of the bulk $(2,2)$ supersymmetry. Thus, any
boundary multiplet will necessarily be a supermultiplet of the
unbroken linear combination. We will loosely refer to these multiplets as
 $(0,2)$ multiplets. While this construction closely parallels
$(0,2)$ constructions for vector bundles in the
GLSM without boundary\cite{zerotwo}, there are also important
differences. For instance, some conditions like the D-term constraint
appear in the low energy analysis by continuity from the bulk. Another
important difference is that one need not impose the vanishing
of the first Chern class of the vector bundle. This is related to the
fact that the fields associated with the vector bundle live on the
boundary of the worldsheet and do not play a role in the bulk R-anomaly
cancellation.  We will
comment on other differences later on in the text, where appropriate.

The strategy in this paper is to construct boundary actions involving
$(0,2)$ multiplets such that together with the bulk action, the 
boundary conditions on the bulk fields preserve the unbroken worldsheet
B-type supersymmetry. This results in the coupling of the boundary multiplets
to the boundary values of the bulk fields. The boundary actions, together
with appropriate bulk-boundary couplings, that we
construct lead in the low-energy limit, in a natural fashion,
to the monad constructions used
for vector bundles. As in \cite{lsmone}, we will
require that our boundary conditions satisfy the following constraints:
\begin{enumerate}
\item[(i)] cancellation of ordinary and supersymmetric variations 
modulo equations of motion; 
\item[(ii)] the set of boundary conditions are closed under the
action of the unbroken supersymmetry; 
\item[(iii)] the boundary conditions 
(especially on the fields in vector multiplets) have
a consistent NLSM limit. The theta term in the bulk requires the
addition of a {\em contact term} which correctly implements the large
volume monodromy action on the vector bundle. This generalises the
contact terms which appeared earlier\cite{lsmone,HIV}. 
\end{enumerate}
Along the way, we
carefully work out the decomposition of bulk multiplets in terms of
boundary $(0,2)$ multiplets. This reorganises the bulk fields
in suitable fashion and motivates the bulk-boundary couplings.

A large class of D-brane configurations can be constructed using the
techniques of this paper. We illustrate this for the large volume
analogues of the $\sum_a l_a=0$  Recknagel-Schomerus (RS) states\cite{RS} at the
Gepner point. The monads associated with these states are precisely
those which appear as mutations of helices associated with the
six-brane. Unlike in most heterotic constructions, the constituents of the 
monads are vector bundles in general. Thus, one can in fact construct
vector bundles which are the cohomology of complexes of line bundles
with length greater than two, in the GLSM.

Our construction can be applied to the cases of D-branes wrapping lower
dimensional cycles. Some of these branes can be obtained from the
transverse intersection of hypersurfaces with the Calabi-Yau manifold.
These fit naturally into the general framework. This is related to the
fact that the cohomology of 
complexes closely related to the monad can give rise to
sheaves. In such situations, one can also reinterpret the result as
a bound state of branes and anti-branes\cite{bpsalgebra}. We also
consider the bound state of two $\sum_a l_a=0$  RS
states to obtain a $\sum_a l_a=1$ RS state.

\section{Background and Notation}
\subsection{The GLSM}

In this section, we will consider the GLSM with $(2,2)$ supersymmetry
on a worldsheet with a boundary which preserves half of the bulk
supersymmetry. We will mostly consider the case of B-type boundary
conditions where the unbroken supersymmetry is given by
$$
\epsilon_-=\eta\epsilon_+ \equiv {\epsilon\over\sqrt2}\quad,
$$
where $\eta=\pm1$.

In order to fix the notations and conventions used in this paper,
we review the Lagrangian and
supersymmetries of the GLSM following \cite{wittenphases}.  We work in
Minkowski space with the metric $(-,+)$.  We are interested in describing
compactifications of string theory with eight supercharges; the worldsheet
conformal field theory must then have $N=(2,2)$ superconformal symmetry. 
We expect that a nonconformal theory with such an infra-red fixed
point should have $N=2$ supersymmetry as well. 

The theory can be obtained by dimensional reduction from
$d=4,N=1$ abelian gauge theory with chiral multiplets.  It contains $s$
$U(1)$ vector multiplets, described by the vector superfields $V^a
(a=1,\cdots, s)$ and $k$ chiral multiplets described by the chiral
superfields $\Phi_i (i= 1,\cdots,k)$. Written in components, the vector
multiplet consists of the vector fields $v^a_{\alpha} (\alpha=0,1)$, the
complex scalar field $\sigma^a$, complex chiral fermions
$\lambda_{\pm}^a$, and the real auxiliary field $D^a$.  The chiral
multiplet consists of a complex scalar $\phi_i$, complex chiral fermions
$\psi_{\pm i}$, and a complex auxiliary scalar field $F_i$. They are
charged under the $U(1)$s with charge $Q_i^a$. In component notation, the
supersymmetry transformations of the vector multiplet are: 
\begin{eqnarray}
\delta v_0^a&=&i \left(\overline{\epsilon}_+ \lambda_+^a 
+\overline{\epsilon}_-\lambda_-^a ~+~\epsilon_+\overline{\lambda}_+^a
~+~\epsilon_-\overline{\lambda}_-^a \right), \nonumber \\
\delta v_1^a&=& i\left( \overline{\epsilon}_+ \lambda_+^a -  
\overline{\epsilon}_- \lambda_-^a+\epsilon_+ \overline{\lambda}_+^a
-\epsilon_- \overline{\lambda}_-^a \right), \nonumber \\
\delta \sigma^a&=&-i \sqrt{2} \overline{\epsilon}_+
\lambda_-^a -i\sqrt{2}\epsilon_-\overline{\lambda}_+^a,\nonumber \\
\delta \overline{\sigma}^a&=&-i \sqrt{2}\epsilon_+\overline{\lambda}_-^a -i
\sqrt{2}\overline{\epsilon}_-\lambda_+^a, \nonumber \\
\delta D^a&=&- \overline{\epsilon}_+(\partial_0 - \partial_1)\lambda_+^a
-\overline
{\epsilon}_-(\partial_0 + \partial_1)\lambda_-^a \\ 
&+&\epsilon_+(\partial_0 -
\partial_1) \overline{\lambda}_+^a +\epsilon_-(\partial_0 + \partial_1)
\overline{\lambda}_-^a, \nonumber \\
\delta \lambda_+^a&=&i\epsilon_+D^a+\sqrt{2}
(\partial_0+\partial_1)\overline{\sigma}^a
\epsilon_- -v^a_{01} \epsilon_+, \nonumber \\
\delta \lambda_-^a&=&i \epsilon_-D^a+\sqrt{2}(\partial_0- \partial_1)\sigma^a
\epsilon_+ +v^a_{01}\epsilon_-, \nonumber \\
\delta \overline{\lambda}_+^a&=&-i \overline{\epsilon}_+D^a
+\sqrt{2}(\partial_0 + \partial_1) \sigma^a\overline{\epsilon}_- 
-v^a_{01}\overline{\epsilon}_+ \nonumber, \\
\delta \overline{\lambda}_-^a&=&-i \overline{\epsilon}_-D^a
+\sqrt{2}(\partial_0 - \partial_1)\overline{\sigma}^a \overline{\epsilon}_+ 
+v_{01}^a\overline{\epsilon}_- \ ,\nonumber
\label{vectortrans}
\end{eqnarray}
where $\epsilon_\pm$ and $\overline{\epsilon}_\pm$ are 
the Grassman parameters for SUSY transformations. 
The transformation rules for the chiral multiplet are:
\begin{eqnarray}
\delta \phi_i&=& \sqrt{2} (\epsilon_+ \psi_{-i}-\epsilon_- \psi_{+i}),
\nonumber \\
\delta \psi_{+i}&=&i \sqrt{2}(D_0 + D_1) \phi_i
\overline{\epsilon}_-+\sqrt{2}\epsilon_+F_i-2Q_i^a \phi_i
\overline{\sigma}^a\overline{\epsilon}_+, \nonumber \\
\delta \psi_{-i}&=&-i \sqrt{2}(D_0 - D_1) \phi_i \overline{\epsilon}_+
+ \sqrt{2} \epsilon_- F_i+2Q_i^a \phi_i \sigma^a\overline{\epsilon}_-,
 \\
\delta F_i&=&-i \sqrt{2} \overline{\epsilon}_+(D_0 - D_1) \psi_{+i}
-i \sqrt{2}\overline{\epsilon}_-(D_0 + D_1)\psi_{-i} \nonumber \\
& & \mbox{} +2Q_i^a(\overline{\epsilon}_+ \overline{\sigma}^a \psi_{-i}
 + \overline{\epsilon}_- \sigma^a\psi_{+i} )
+2iQ_i^a\phi_i(\overline{\epsilon}_- \overline{\lambda}_+^a- 
\overline{\epsilon}_+ \overline{\lambda}_-^a)
\label{chiraltrans}
\end{eqnarray}

The supersymmetric bulk action can be written as a sum of four terms, 
\begin{equation}
S=S_{ch}+S_{gauge}+S_{W}+S_{r,\theta}
\label{bulkaction}
\end{equation}
The terms on the right hand side are, respectively: the kinetic term for
the chiral superfields; the kinetic terms for the vector superfields; the
superpotential interaction; and the Fayet-Iliopoulos and theta terms. 
$S_{ch}$ is: 
\begin{eqnarray}
S_{ch}&=&\sum_i\int d^2x \left\{ 
- D_{\alpha}\overline{\phi}_i D^{\alpha} \phi_i
+i \overline{\psi}_{-i}( \stackrel{\leftrightarrow}{D_0}
+\stackrel{\leftrightarrow}{D_1})\psi_{-i} \right.
+i \overline{\psi}_{+i}
( \stackrel{\leftrightarrow}{D_0}
-\stackrel{\leftrightarrow}{D_1})
\psi_{+i} \nonumber \\
& & \mbox{}+~|F_i|^2 ~-~2\sum_a\overline{\sigma}^a\sigma^a 
(Q_i^a)^2\overline{\phi}_i
\phi_i~-~\sqrt{2}\sum_a Q_i^a(\overline{\sigma}^a\overline{\psi}_{+i}\psi_{-i}~
+~\sigma^a
\overline{\psi}_{-i}\psi_{+i}) \nonumber \\ 
& & \mbox{}~+~D^aQ_i^a\overline{\phi}_i\phi_i 
-i\sqrt{2}\sum_aQ_i^a\overline{\phi}_i(\psi_{-i}
\lambda_+^a~-~\psi_{+i} \lambda_-^a) \nonumber\\
& & \left. \mbox{}~-~i \sqrt{2}Q_i^a\phi_i
(\overline{\lambda}_-^a
\overline{\psi}_{+i}~-~\overline{\lambda}_+^a\overline{\psi}_{-i} ) 
\right\}
\label{kechiral}
\end{eqnarray}
where
\begin{equation}
	A \stackrel{\leftrightarrow}{D_i} B \equiv 
	{1\over2} (A D_i B - (D_iA)B)\ .
\label{symder}
\end{equation}
This symmetrized form of the fermion kinetic term
is Hermitian in the presence of a boundary.
Meanwhile, $S_{gauge}$ is:
\begin{eqnarray}
S_{gauge}&=&\sum_a{1 \over e_a^2}\int d^2x\left\{{1 \over 2}(v_{01}^a)^2 + 
{1 \over 2}(D^a)^2 - \partial_{\alpha}\sigma^a \partial^{\alpha} 
\overline{\sigma}^a \right.  \nonumber \\
&&+  \left. i \overline{\lambda}_+^a
(\stackrel{\leftrightarrow}{\partial_0}-
\stackrel{\leftrightarrow}{\partial_1})
\lambda_+^a +i \overline{\lambda}_-^a
(\stackrel{\leftrightarrow}{\partial_0}+
\stackrel{\leftrightarrow}{\partial_1})
\lambda_-^a \right\}
\label{kevector}
\end{eqnarray}
The superpotential term is:
\begin{equation}
S_W=-\int d^2x \left( F_i {\partial W \over \partial
\phi_i}~+~{\partial^2 W \over \partial \phi_i\partial\phi_j} \psi_{-i}
\psi_{+j}~+~{\overline F}_i
{\partial \overline{W} \over \partial \overline{\phi}_i}~-~{\partial^2
\overline{W}
\over
\partial\overline{\phi}_i\partial\overline{\phi}_j}
\overline{\psi}_{-i}\overline{\psi}_{+j}
\right)\ .
\label{superpot}
\end{equation}
Finally, the Fayet-Iliopoulos D-term and theta term are:
\begin{equation}
S_{r, \theta}=-r_a\int d^2x D^a+{\theta_a\over 2 \pi} \int d^2 x
v_{01}^a\ .
\label{fid}
\end{equation}

The bosonic potential energy is given by
\begin{equation}
U=\sum_i\left|
F_i\right|^2+
\sum_a\left(
\frac{D^a}{2e^2}+2|\sigma^a|^2
\sum_i{Q^a_i}^2|\phi_i|^2\right)\ .
\label{bosonicpe1}
\end{equation}
The auxiliary fields $D$ and $F_i$ can be eliminated by their
equations of motion:
\begin{eqnarray}
D^a&=&-e^2\left(\sum_iQ^a_i|\phi_i|^2-r^a\right)\nonumber \\
F_i^*&=&\frac{\partial W} {\partial\phi_i}\ ,
\label{dandf}
\end{eqnarray}

In this paper, we will be mostly considering the case of a single $U(1)$ gauge
field though the generalisation to many $U(1)$'s is obvious. 
Consider $(n+1)$ chiral superfields $\Phi_i$ of positive charge $Q_i$
($i=1,\ldots,{n+1}$)
and one superfield $\Phi_0\equiv P$ of charge $Q_0=Q_p=-\sum_{i\neq0}Q_i$. 
In the absence
of a superpotential, in the NLSM limit, the target space is a
non-compact Calabi-Yau manifold given by
the total space of a line bundle ${\cal O}(Q_p)$ over
the weighted projective space $\BP^{Q_1,\ldots,Q_{n+1}}$. 
A quasi-homogeneous superpotential of $W=PG(\Phi)$, where $G$ is a
degree $|Q_p|$ polynomial (satisfying certain transversality
conditions\cite{wittenphases}) involving chiral fields other than $P$ gives
rise to a compact Calabi-Yau manifold which is the  hypersurface $G=0$
in the weighted projective space.

\subsection{Bulk and Boundary Supermultiplets}
\label{secdecomp}

When A- or B-type boundary conditions are imposed, one half
of the bulk $(2,2)$ supersymmetry is broken. Let $Q_{\pm}$ and
$\overline{Q}_{\pm}$ be the generators of the $(2,2)$ supersymmetry
algebra. The unbroken generators are given by the linear combinations
$$ Q \equiv {1\over\sqrt2} (Q_- - \eta \overline{Q}_+) $$  for A-type
boundary conditions and by $$Q \equiv {1\over\sqrt2} (Q_- + \eta Q_+)$$ 
for B-type boundary conditions. They
satisfy the $0+1$ dimensional supersymmetry algebra
$$
\{Q,\overline{Q}\} = 2 P_0 \quad,
$$
where $P_0$ is the generator of translations along $x^0$ and the
other anticommutators are vanishing.  

\subsubsection{Boundary Superspace description}

In superspace with coordinates $(x^0,\theta,\overline{\theta})$,
the supersymmetry generators have the following
representation\footnote{The Grassmann parameters $\theta$ and
$\overline{\theta}$ are related to the bulk Grassmann parameters.
For example, $\theta^-=\eta\theta^+ =\sqrt2\theta$ on the boundary
for B-type boundary conditions. 
This can also be viewed as the definition of the boundary in
superspace.}:
\begin{equation}
Q= {\partial\over{\partial\theta}} + i \overline{\theta}
\partial_0 \quad,\quad
\overline{Q}= -{\partial\over{\partial\overline{\theta}}} - i \theta
\partial_0 \quad,
\end{equation}
where $\partial_0\equiv \partial/\partial x^0$.
The superderivatives which commute with the supersymmetry generators
are
\begin{equation}
D= {\partial\over{\partial\theta}} - i \overline{\theta}
\partial_0 \quad,\quad
\overline{D}= -{\partial\over{\partial\overline{\theta}}} + i \theta
\partial_0 \quad.
\end{equation}
\bigskip
\begin{flushleft}
{\underline{\bf The Gauge Multiplet}}
\end{flushleft}

Gauge fields are introduced in superspace by means of gauge covariant
derivatives ${\cal D}$, $\overline{\cal D}$ and ${\cal D}_0$
satisfying the constraints
\begin{eqnarray}
{\cal D}^2 &=& \overline{\cal D}^2 =0 \nonumber \\
\{{\cal D}, \overline{\cal D} \} &=& 2i {\cal D}_0\quad.
\end{eqnarray}
One can solve for the above constraints by introducing a Lie algebra
valued real superfield $V$ such that
\begin{eqnarray}
{\cal D} &=& e^{-V}\ D\ e^{V} \nonumber \\
\overline{\cal D} &=& e^{V}\ \overline{D}\ e^{-V}\quad.
\end{eqnarray}
In the analogue of the Wess-Zumino gauge, $V=\theta \overline{\theta}
v_0$ with
\begin{eqnarray}
{\cal D}_0 &=& \partial_0 + i v_0 \nonumber \\
{\cal D} &=& {\partial\over{\partial\theta}} - i \overline{\theta}
{\cal D}_0 \\
\overline{\cal D}&=& -{\partial\over{\partial\overline{\theta}}} + i \theta
{\cal D}_0 \nonumber
\end{eqnarray}
In the Wess-Zumino gauge, $\delta v_0 =0$ i.e., the gauge field is
invariant under supersymmetry transformations (such that the
Wess-Zumino gauge is preserved). Further, there is no kinetic energy
term for the gauge field since we are in $0+1$ dimensions.

\begin{flushleft}
\underline{\bf Chiral Multiplets}
\end{flushleft}

Chiral multiplets (with $U(1)$ charge $Q$) satisfy
$$
\overline{\cal D} \Phi =0
$$
and have the following component expansion
\begin{equation}
\Phi = \phi + \sqrt2\theta \tau -i \theta \overline{\theta}
D_0 \phi\quad,
\end{equation}
where $D_0 \phi = (\partial_0 + iQ v_0 ) \phi$ is the usual covariant
derivative.
The supersymmetry transformation of the component fields are
\begin{eqnarray}
\delta\phi &=&  \sqrt2\epsilon \tau \\
\delta \tau &=& -\sqrt2 i \overline{\epsilon} D_0 \phi
\end{eqnarray} 
The kinetic energy term for chiral superfields is
given by
\begin{eqnarray}
S&=&-{i\over2}\int dx^0 d^2\theta \left(
\overline{\Phi} D_0 \Phi \right) \\
&=& \int dx^0 \left( |D_0\phi|^2 +i \overline{\tau}D_0 \tau\right)
\end{eqnarray}

We will also need to consider fermi multiplets $\Pi$ with
components $(\pi,l)$ satisfying the constraint
\begin{equation}
\overline{\cal D} \Pi = \sqrt2 E(\Phi)\quad,
\end{equation}
where $E(\phi)$ is a function of chiral superfields $\Phi_i$. 
The component expansion of the superfield $\Pi$ is
\begin{equation}
\Pi = \pi - \sqrt2 \theta l - \overline{\theta}\sqrt2  E(\phi)
+ \theta \overline{\theta} \left(-i D_0 \pi
+2 \tau_i {{\partial E}\over{\partial\phi_i}}\right)\\ \quad.
\end{equation}
When $E=0$, this reduces to the expansion of a chiral superfield.
The supersymmetry transformation of the fields in the Fermi multiplet are
\begin{eqnarray}
\delta\pi &=&-\sqrt2 \epsilon l -\sqrt2 \overline{\epsilon}E(\phi)\\
\delta l &=&  i\sqrt2 \overline{\epsilon} D_0 \pi 
-\sqrt2 \overline{\epsilon}\tau_i {{\partial E}\over{\partial \phi_i}}
\end{eqnarray}
Consider Fermi superfields satisfying
\begin{equation}
\overline{\cal D} \Pi_a = E_a(\Phi)
\end{equation}
where $E_a$ are chiral superfields. 
The supersymmetric action for Fermi multiplets is
\begin{equation}
S_F = -{1\over2}\int dx^0  d^2\theta\ \overline{\Pi}_a \Pi_a
\end{equation}
The component expansion of the above action is
\begin{equation}
S_F=\int dx^0 \left( i\overline{\pi}_aD_0 \pi_a + |l_a|^2 -
|E_a(\phi)|^2 - \overline{\pi}_a 
{{\partial E_a}\over{\partial \phi_i}}\tau_i - 
{{\partial \overline{E}_a}\over{\partial \overline{\phi}_i}}
\overline{\tau}_i \pi_a \right)
\end{equation}

Further, let $J^a(\Phi)$ be chiral
superfields satisfying the condition
\begin{equation}
E_a J^a =0\quad.
\end{equation}
One can then introduce a boundary superpotential of the form
\begin{equation}
S_J=-{1\over\sqrt2} \int dx^0 d\theta (\Pi_a J^a)|_{\overline{\theta}=0} -
{\rm h.c.}
\end{equation}
The component expansion is
\begin{equation}
S_J = -\int dx^0 \left( l_a J^a(\phi) + 
\pi_a {{\partial J^a}\over{\partial \phi_i}} \tau_i  
+ \overline{l}_a \overline{J}^a(\overline{\phi}) 
+ {{\partial \overline{J}^a}\over{\partial
\overline{\phi}_i}} \overline{\tau}_i  \overline{\pi}_a
\right) 
\end{equation}

On eliminating the auxiliary fields $l_a$ using their equations of
motion, we obtain the boundary Lagrangian
\begin{eqnarray}
\lefteqn{S_F + S_J = \int dx^0 \left(
i\overline{\pi}_a\widetilde{D}_0 \pi_a - |J^a(\phi)|^2 -
|E_a(\phi)|^2 \phantom{1\over2}\right.} \hspace{1in}\nonumber \\
&&- \overline{\pi}_a
{{\partial E_a}\over{\partial \phi_i}}\tau_i 
 \left.
- {{\partial \overline{E}_a}\over{\partial \overline{\phi}_i}}
\overline{\tau}_i \pi_a
- \pi_a {{\partial J^a}\over{\partial \phi_i}} \tau_i
- {{\partial \overline{J}^a}\over{\partial
\overline{\phi}_i}} \overline{\tau}_i  \overline{\pi}_a
\right)
\end{eqnarray}

\subsubsection{Decomposition of bulk multiplets: B-type}

We now study how  $(2,2)$ multiplets in the bulk
decompose as boundary multiplets. We will consider the case of B-type
boundary conditions. The boundary supersymmetry parameter $\epsilon$
is related to the bulk parameters by $$\epsilon=\sqrt2 \epsilon_-
=\sqrt2 \eta \epsilon_+\quad .$$ In addition, the 
boundary supercoordinates are related to the bulk ones by
$$
\sqrt2\theta=\theta^-=\eta\theta^+ 
$$
\begin{flushleft}
\underline{\bf $(2,2)$ Vector Multiplets}
\end{flushleft}

The vector multiplet decomposes into the following combinations:
\begin{enumerate}
\item  $\widetilde{v}_0=v_0+\eta{{(\sigma+ \overline{\sigma})}\over\sqrt2}$
transforms as a singlet under the unbroken supersymmetry 
and is the $(0,2)$ vector multiplet in the Wess-Zumino gauge.
\item The bulk twisted chiral superfield $\Sigma$ becomes an
unconstrained  complex $(0,2)$ superfield with the following
expansion
\begin{equation}
\Sigma = \sigma -2i\eta\theta \overline{\lambda}_+ -2i\overline{\theta}
\lambda_- + 2 \sqrt2\theta \overline{\theta}\eta (\widetilde{D}
-i\widetilde{v}_{01}) 
\end{equation}
where $\widetilde{D}\equiv D +i\eta
\partial_1{{\sigma-\overline{\sigma}}\over\sqrt2}$ and
$\widetilde{v}_{01} = \partial_0 v_1  - \partial_1 \widetilde{v}_0$.
\item It is also useful to note that the combination
$\widetilde{v}_1\equiv
v_1-\eta {{(\sigma-\overline{\sigma})}\over\sqrt2}$, $(\lambda_+ -\eta
\lambda_-)$ form a chiral superfield (we will call it $V_1$) if we choose
$\partial_1 \widetilde{v}_0=0$ on the boundary. 
\end{enumerate}

\begin{flushleft}
\underline{\bf $(2,2)$ Chiral Multiplets}
\end{flushleft}

A $(2,2)$ chiral superfield $\Phi$ decomposes into two $(0,2)$
multiplets: 
\begin{enumerate}
\item A chiral multiplet $\Phi'$ with components $(\phi,\tau)$ 
with $\tau\equiv (\psi_- - \psi_+)/\sqrt2$. The boundary Lagrangian
for chiral multiplets $\Phi'_i$ given by
\begin{equation}
S_{\rm pert} = {1\over2}\int dx^0 d^2\theta F^{i\overline{\jmath}} 
\Phi'_i \overline{\Phi}'_j
\end{equation}
corresponds to turning on a constant field strength $F^{i\overline{\jmath}}$
in the worldvolume of the brane. It has the following
component expansion:
\begin{equation}
S_{\rm pert} = \int dx^0  F^{i\overline{\jmath}}
\left[ {i\over2} (\phi_i \widetilde{D}_0 \overline{\phi}_j -
\overline{\phi}_j \widetilde{D}_0 \phi_i) + \tau_i \overline{\tau}_j \right]
\end{equation}
\item A fermi multiplet
$\Xi$ with components $(\xi\equiv {{\psi_-+\psi_+}\over\sqrt2},-F)$ satisfying 
$$ \overline{\cal D} \Xi =-i\sqrt2 \widetilde{D}_1 \Phi' \quad. $$
where $\widetilde{D}_1\Phi'\equiv (\partial_1 +i Q V_1)\Phi'$. On the
boundary, we treat $\partial_1\Phi'$ as an independent $(0,2)$ chiral
superfield with components $(\partial_1\phi,\partial_1\tau)$.
\end{enumerate}

\section{Holomorphic vector bundles from complexes}

In anticipation of the fact that in the low-energy limit, we expect
our construction to reduce to the construction of vector bundles
as the cohomology of complexes such as monads(which are complexes of
length two),
we now discuss in this section
some relevant aspects of the monad construction followed by the a
discussion of the cases where the complexes  have length greater than
two.
We also discus in some detail
the monads associated with the $\sum_al_a=0$ RS states for the case of
Calabi-Yau manifolds given by hypersurfaces in weighted projective
space.

Monads are a construction originally due to Horrocks and used extensively by
Beilinson for constructing holomorphic vector
bundles on $\BP^n$. (For a readable introduction to this subject, see
\cite{okonek}.)
The basic idea is to consider the following complex 
(monad) of holomorphic vector bundles  $A$, $B$ and $C$ 
\begin{equation}
0\rightarrow A \stackrel{a}{\rightarrow} B \stackrel{b}{\rightarrow}
C \rightarrow 0\quad,
\end{equation}
which is exact at $A$ (equivalently, the map $a$ is injective)
and $C$ (equivalently, the map $b$ is surjective)
such that Im$(a)\subset B$ and $b\cdot a=0$.
The holomorphic
vector bundle $$E={\rm ker}\ b/{\rm Im}\ a$$ is the cohomology of the
monad.  Some topological properties of $E$ are
\begin{eqnarray}
{\rm rk}\ E &=& {\rm rk}\ B - {\rm rk}\ A - {\rm rk}\ C \\
{\rm ch}(E) &=& {\rm ch}(B) - {\rm ch}(A) -{\rm ch}(C)
\end{eqnarray}

We will also consider the exact sequences
$$
0\rightarrow A \stackrel{a}{\rightarrow} B \rightarrow
E \rightarrow 0\quad,
$$
(with ${\rm rk}\ E = {\rm rk}\ B - {\rm rk}\ A$ and
${\rm ch}(E) = {\rm ch}(B) - {\rm ch}(A)$)
and 
$$
0\rightarrow E \rightarrow B \stackrel{b}{\rightarrow}
C \rightarrow 0\quad,
$$
(with ${\rm rk}\ E = {\rm rk}\ B - {\rm rk}\ C$ and
${\rm ch}(E) = {\rm ch}(B) - {\rm ch}(C)$)
in order to construct holomorphic vector bundles.

The field theoretic construction of these vector bundles is as
follows\cite{wittenphases}.
Consider fermions $\pi_a$ ($a=1,\ldots,{\rm rk}\ B$) and $\kappa_i$
($i=1,\ldots,{\rm rk}\ A$)
which are sections of $B$ and $A$ respectively.  The map $a$ 
(represented by  $E_a^i(\phi)$) is realised as the gauge invariance
$$
\pi_a \sim \pi_a + E^i_a(\phi) \kappa_i \quad.
$$
This gauge invariance is fixed by the gauge choice
\begin{equation}
\overline{E}_a^i(\phi) \pi_a =0
\label{conda}
\end{equation}
The map $b$ (represented by $J^a_m(\phi)$) imposes the holomorphic
constraints
\begin{equation}
J^a_m(\phi)\pi_a =0\quad m=1,\ldots,{\rm rk}\ C
\label{condb}
\end{equation}
The remaining fermions, i.e., those not set to zero by (\ref{conda}) and
(\ref{condb}), are sections of the holomorphic vector bundle $E$ given
by the monad construction.  

Given a vector bundle $E$ as the cohomology of a monad with constituents
$A$, $B$, $C$ as above, one can verify that the vector bundle
$E(n)=E\otimes {\cal O}(n)$ is given by the cohomology of the following
monad
\begin{equation}
0\rightarrow A(n) \stackrel{a}{\rightarrow} B(n) \stackrel{b}{\rightarrow}
C(n) \rightarrow 0\quad,
\end{equation}
In the field theoretic construction that we will pursue, the operation
of tensoring with ${\cal O}(n)$ corresponds to shifting the charges of
all the fermions by $n$ units.

One may encounter situations where the map $a$ is not injective on
some submanifold $\Sigma$. Then, one obtains a sheaf rather than a vector bundle
whose singularity set is $\Sigma$. A simple example which illustrates
this is to consider a single fermion which is a section of ${\cal O}$.
We choose $E=\phi_1$. Then, the condition
$$
\overline{\phi}_1 \pi =0
$$
sets $\pi=0$ on all points except the hyperplane $\phi_1=0$. Thus, the
fermion $\pi$ is non-vanishing on the hyperplane and is a section
of the sheaf of functions with support on the hyperplane. We will see
that this is useful in constructing lower dimensional branes i.e.,
D-branes wrapping some holomorphic sub-cycle of the Calabi-Yau manifold
rather than the whole Calabi-Yau manifold.

\subsection{Vector bundles for $\BP^n$}
\label{secvbpn}

For the case of $\BP^n$, we would like to construct the bundles
corresponding to the $l=0$ orbit. The homogeneous coordinates on $\BP^n$
are given by $(\phi_1,\ldots,\phi_{n+1})$.
Introduce $\left( {}^{n+1}_{m+1}\right)$ fermions
$\pi_{[i_1,\cdots,i_{m+1}]}$ $(i_1,\cdots,i_{m+1}=1,\ldots,n+1)$
subject to the conditions
\begin{equation}
\overline{E}_{[i_1,\cdots,i_{m+1}]}^{[j_1,\cdots,j_m]}
\pi_{[i_1,\cdots,i_{m+1}]} =0\quad,
\label{econd}
\end{equation}
where 
$$
E_{[i_1,\cdots,i_{m+1}]}^{[j_1,\cdots,j_m]} = {1\over{(m+1)!}}
\sum_{\rm all\ perms} (-)^p
\phi_{i_{p(1)}} \delta_{i_{p(2)}}^{j_1}\cdots\delta_{i_{p(m+1)}}^{j_m}\quad.
$$
For example, when $m=1$,
$E_{ij}^k = (\phi_i \delta_j^k - \phi_j \delta_i^k)/2$.
One can verify that the following is true
$$
\phi_{j_1} E_{[i_1,\cdots,i_{m+1}]}^{[j_1,\cdots,j_m]} =0
$$
Thus, the number of independent $E$'s are $\left(
{}^{n}_{m}\right)$ and hence the number of fermions remaining after
we impose the conditions (\ref{econd}) is equal to
$\left({}^{~n}_{m+1}\right)$. 
The remaining fermions transform as a section of $\BT^{m+1}(-m-1)$ 
-- the $(m+1)$-th
exterior power of the tangent bundle tensored by ${\cal O}(-m-1)$. 
When $m=0$, this is seen by considering the Euler sequence
\begin{equation}
0\rightarrow {\cal O}(-1)\rightarrow {\cal O}^{\oplus (n+1)}\rightarrow
\BT_{\BP^n}(-1)\rightarrow 0
\label{euler}
\end{equation}
The general case is given by the exact sequence (derivable from the 
Euler sequence)
\begin{equation}
0\rightarrow \BT^{p-1}_{\BP^n}(-p)
\stackrel{E_i}{\rightarrow} {\cal O}^{\oplus ({}^{n+1}_{~p})}\rightarrow
\BT^p_{\BP^n}(-p)\rightarrow 0
\end{equation}
where $\BT^p_{\BP^n}\equiv\wedge^p \BT\BP^n$ is the $p$-th exterior power
of the tangent bundle to $\BP^n$. Now, consider the exact sequence 
which is dual to the above one
\begin{equation}
0\rightarrow \Omega^{p}_{\BP^n}(p)
\rightarrow {\cal O}^{\oplus ({}^{n+1}_{~p})}\rightarrow
\Omega^{p-1}_{\BP^n}(p)\rightarrow 0
\label{dualeuler}
\end{equation}
where $\Omega_{\BP^n}$ is the cotangent bundle to $\BP^n$. This bundle
is constructed by choosing conditions imposed by $J$'s instead of the
$E$'s. For example, to obtain the cotangent bundle, we consider $(n+1)$
fermions $\pi^i$ subject to the (holomorphic) constraint
$$
\phi_i \pi^i =0 \quad.
$$
Thus, given a holomorphic vector bundle $E$ as the cohomology of a
monad, its dual $E^\ast$  is given by a monad where
the gauge conditions are exchanged with the constraints.  
\begin{center}
\begin{tabular}{|ccc|}  \hline
Vector bundle $E$ &$\leftrightarrow$&  Its dual $E^\ast$ \\ \hline
Gauge conditions &$\leftrightarrow$& Holomorphic constraints \\
Holomorphic constraints &$\leftrightarrow$& 
Gauge conditions \\ \hline
\end{tabular}
\end{center}
As we will see, in the field theoretic realisation,
this has a very simple relation.

\subsection{Long sequences and $\Omega^2(2)$}

As we have seen in the previous subsection, the monad construction
seems to lead us to sequences of length two (i.e., those involving
three bundles). Further, monads involving vector bundles may be
represented by longer sequences involving only line bundles. This is
best illustrated by considering the case of $\Omega^2(2)$. 
By combining the monad used for $\Omega^1(1)$, one can obtain
the following sequence of length three which leads to $\Omega^2(2)$.
\begin{equation}
0\rightarrow \Omega^{2}(2)
\rightarrow {\cal O}^{\oplus 10}\rightarrow
{\cal O}^{\oplus 5}(1)\rightarrow
{\cal O}(2)\rightarrow 0 
\end{equation}
As we will see, in the field theoretic construction, it is more natural
for the fermions to be sections of direct sums of line bundles and 
thus, we will need to implement a sequence of length three in order to
obtain $\Omega^2(2)$. We will postpone the precise details to the next
section.

More generally, from the work of Beilinson\cite{beilinson},
all coherent sheaves on
$\BP^n$ arise from sequences of length less than or equal to $n$.
Thus, it is of interest to be able to deal with sequences which are of
length greater than two.

\subsection{Vector bundles for weighted projective spaces}
\label{secvbwpn}

Weighted projective spaces are typically singular. We will be interested
in situations where the Calabi-Yau is embedded as an hypersurface in
some weighted projective space. There are two possible scenarios in this
context: (i) the hypersurface does not inherit any of the singularities of 
the ambient weighted projective space; (ii) the weighted projective
space does inherit some of the singularities of the ambient projective
space. For case (ii) we assume that one can make the Calabi-Yau hypersurface
non-singular by blowing up the singularities. This process converts
case (ii) to that of case (i).

In a recent paper\cite{helices}, we have constructed  rigid sheaves
on weighted projective spaces by suitably generalising the mutations of
helices on these spaces (see also \cite{mayrtomas}). 
This construction has the advantage of 
being adaptable to the GLSM construction of vector bundles pursued in
this paper. We will illustrate this for the case of of a degree
six hypersurface in $\BP^{1,1,1,1,2}$. Let $\phi_i$ for $i=1,\ldots,5$
be the quasi-homogeneous coordinates of the weighted projective space
with $\phi_5$ having weight two.
The $\sum_a l_a=0$ bundles
$S_i$, $i=1,\ldots,6$ are related to each other under the quantum
$\BZ_6$ symmetry. $S_1={\cal O}$.

The bundle $-S_2$ (the minus sign reflects the K-theory class)
is defined by the left mutation
\begin{equation}
0\rightarrow (-S_2) \rightarrow  {\cal O}^{\oplus 4}
\stackrel{J}{\rightarrow} {\cal O}(1) \rightarrow 0
\end{equation}
where $J^i=\phi_i$ for $i=1,2,3,4$. This sequence is similar to the
Euler sequence associated with $\BP^3$ (with homogeneous coordinates
$\phi_1,\ldots,\phi_4$) and is a bundle of rank three. Thus $(-S_2)$
is closely related to the cotangent bundle of the $\BP^3$ (i.e.,
$\Omega(1)$) in the chart $\phi_5=0$.
The next bundle is given by the exact sequence
\begin{equation}
0\rightarrow S_3
\rightarrow {\cal O}^{\oplus 5}\rightarrow
(-S_2)\otimes{\cal O}(1)\rightarrow 0
\end{equation}
We will instead consider the sequence (similar to the one associated
with $\Omega^2(2)$ for $\BP^3$)
\begin{equation}
0\rightarrow S'_3 \rightarrow {\cal O}^{\oplus 6} \rightarrow 
(-S_2)\otimes{\cal O}(1)\rightarrow 0
\end{equation}
We consider six fermions $\pi^{[ij]}$ ($i,j=1,2,3,4$) subject to
the holomorphic constraints $J^k_{ij} \pi^{[ij]}=0$, 
where $J^k_{ij}= {1\over2}(\phi_i \delta_j^k - \phi_j\delta_i^k)$.
$S'_3$ is thus a rank three vector bundle.
The bundle of interest $S_3$ is in the K-theory class $(S'_3-{\cal O})$
as can be verified by comparing the Chern classes\cite{helices}.
The other three bundles are given by Serre duality:
$S_i \simeq S_{7-i}^\ast \otimes {\cal O}(-1)$.  The  restriction of
these bundles to the degree six hypersurface gives rise to bundles
$V_i$ which can be shown to reproduce the charges associated with
the $\sum_a l_a=0$ D-branes\cite{helices}.

\subsection{Vector bundles for hypersurfaces in $\BP^n$}

We will be interested in vector bundles for Calabi-Yau manifolds given by
hypersurfaces or transverse intersection of hypersurfaces in $\BP^n$. 
For simplicity, we will only consider the case of a hypersurface $M$ given by
a homogeneous polynomial $G(\phi)$ of degree $Q_p$.  
One simple class of vector bundles are given by
the restriction $E|_M$ of vector bundles on $\BP^n$ to the hypersurface $X$.
Thus, the $L=0$ bundles are restrictions to bundles on $M$. 

The tangent bundle $\BT_M$ is obtained by considering $(n+1)$ fermions
subject to the conditions given by $E_i=\phi_i$ and $J^i=(\partial
G/\partial \phi_i)$. The tangent bundle is thus given by the monad
$$
0\rightarrow {\cal O}\stackrel{\otimes E_i}{\longrightarrow}
 {\cal O}(1)^{\oplus n+1}\stackrel{\otimes J^i}{\longrightarrow}
{\cal O}(Q_p)\rightarrow 0
$$

\section{\large The monad construction in the GLSM with boundary}

In addition to the standard Lagrangian for the bulk GLSM, on the boundary,
we introduce fermi multiplets $\Pi_a$ satisfying\footnote{This
constraint is interpreted as the gauge fixing of a fermionic gauge
invariance. In this paper, we always work in this gauge-fixed formulation.
For details, see \cite{wittenphases,zerotwo}.}
\begin{equation}
\overline{\cal D} \Pi_a = \sqrt2 \Sigma' E_a(\Phi')\quad,
\end{equation}
where $\Phi'$ is the bulk $(0,2)$ chiral multiplet and $\Sigma'$
is a boundary chiral multiplet with components $(\varsigma,\beta)$. 
We will also introduce a superpotential coupling $P' J^a(\Phi')$, where
$P'$ is a boundary chiral multiplet with components $(p',\gamma)$.
The boundary Lagrangian is 
\begin{eqnarray}
S^{1}_{\rm bdry} &=& \int dx^0 \left(
i\overline{\pi}_a\widetilde{D}_0 \pi_a - |p'|^2|J^a(\phi)|^2 -
|\varsigma|^2|E_a(\phi)|^2 \right.  \nonumber \\
&&- E_a \overline{\pi}_a \beta
- \overline{E}_a \overline{\beta} \pi_a 
- a_1 \overline{\pi}_a
\varsigma{{\partial E_a}\over{\partial \phi_i}}\tau_i 
- a_1 \overline{\varsigma}
{{\partial \overline{E}_a}\over{\partial \overline{\phi}_i}}
\overline{\tau}_i \pi_a  \\
&&- \overline{J}^a \overline{\gamma}\overline{\pi}_a 
- J^a \pi_a \gamma
\left.
-a_2 \pi_a p' {{\partial J^a}\over{\partial \phi_i}} \tau_i 
-a_2 \overline{p}'{{\partial \overline{J}^a}\over{\partial
\overline{\phi}_i}} \overline{\tau}_i  \overline{\pi}_a
\right) \nonumber
\end{eqnarray}
where $a_1$ and $a_2$ are two real constants which are determined by the
condition that the boundary terms in the ordinary and supersymmetric
variation of the Lagrangian
$S_{\rm bulk} + S_{\rm bdry}$  vanish. The choice $a_1=a_2=1$
gives the standard $(0,2)$ Lagrangian as discussed in sec.
\ref{secdecomp}.
Further, the covariant derivative given by
$\widetilde{D}_0 = \partial_0 +iQ \widetilde{v}_0$ involves the combination
$\widetilde{v}_0=v_0 + \eta {{\sigma+\overline{\sigma}}\over\sqrt2}$
as is appropriate from the boundary decomposition of
the bulk vector multiplet as discussed in sec \ref{secdecomp}. 
In the heterotic $(0,2)$ constructions, the
the fermions $\pi_a$ are chosen to have charge
to equal the degree of $E_a$. We will  initially choose the fermions to
be charge neutral and as a consequence, the $\Sigma'$ field has charge
$-Q_a$ (where $Q_a$ is the degree of the $E_a$). However, in what
follows, we will always use the covariant derivatives on the boundary
fermions since we will eventually discuss the cases when they are
charged.

We shall choose a first-order kinetic term for the $\Sigma'$ and $P'$
superfields associated with the gauge-invariance  and holomorphic
constraints. This is a departure from the standard $(0,2)$ construction
in the context of the heterotic string. Our motivation is two-fold: 
\begin{itemize}
\item[(i)] A first-order action is (almost) unavoidable for the case of
complexes of length greater than two. (see sec. \ref{seclong})
\item[(ii)] The large-volume monodromy associated with the bundles is obtained
in a simple manner. (see sec. \ref{seclvmon})
\end{itemize}
The action that we choose is
\begin{equation}
S^2_{\rm bdry} = \int dx^0 \left(
{i\over2} (\varsigma \widetilde{D}_0 \overline{\varsigma} -
\overline{\varsigma} \widetilde{D}_0 \varsigma) + \beta \overline{\beta}
+ {i\over2} (p' \widetilde{D}_0 \overline{p}' -
\overline{p}' \widetilde{D}_0 p') + \gamma \overline{\gamma}
\right)
\end{equation}

\subsection{$\theta=0$ and the low-energy limit}

We will now consider the low-energy limit of this field theory when the
bulk is its geometric phase (which can be obtained by the usual 
$e^2r\rightarrow\infty$ limit). The coupling $\overline{E}_a
\overline{\beta} \pi_a$  is a mass-term for the fermionic combination
$\overline{E}_a \pi_a$ and hence one obtains $\overline{E}_a \pi_a=0$
at energy scales much smaller than this mass scale.
Similarly,  one obtains the holomorphic
constraint $J^a \pi_a=0$  at low-energies from the mass-term
$J^a\pi_a \gamma$. Further, when $\sum_a |E_a|^2$ is non-vanishing
(which we assume for now), the ground state condition requires
$\varsigma=0$. Similarly, $p'=0$ at low energies. These  arguments
parallel those for the $(0,2)$ constructions for the heterotic
string\cite{zerotwo}.

The corresponding analysis for the bulk fields is standard and
we will not repeat them here\cite{wittenphases}.
We recall that fields in the vector multiplet
become Lagrange multipliers enforcing
constraints (These constraints have been explicitly presented in
ref. \cite{lsmone}). We shall quote some of the relevant ones
\begin{eqnarray}
{{\sigma - \overline{\sigma}}\over{\sqrt2}} =
{1\over{2K[\phi]}} \sum_i Q_i\left[ \overline{\tau}_i \xi_i -
\overline{\xi}_i \tau_i \right] \\
{{\sigma + \overline{\sigma}}\over{\sqrt2}} = -
{1\over{2K[\phi]}} \sum_i Q_i\left[ \overline{\xi}_i \xi_i -
\overline{\tau}_i \tau_i \right] \\
\widetilde{v}_0 = {1\over{K[\phi]}} \sum_i Q_i\left[ {i\over2}
(\overline{\phi}_i\partial_0 \phi_i -\phi_i \partial_0
\overline{\phi}_i) + \overline{\tau}_i \tau_i \right] \label{stgaugefield}
\end{eqnarray}
where $K[\phi]=\sum_i Q_i^2 |\phi_i|^2$.

We will follow the strategy pursued in \cite{lsmone} in obtaining
boundary conditions. We  first
obtain boundary conditions for the fields in the matter multiplets in
the NLSM limit as well as equations of motion for the boundary fermi
multiplets such that all boundary terms which appear on the ordinary and
supersymmetric variations (involving these fields) of the action vanish. 
Next, one obtains conditions on the fields in the vector
multiplet and the equations of motion for the $\Sigma'$ and $P'$ fields. 
Finally, these conditions are lifted to the GLSM by
requiring that the boundary terms of order $1/e^2$ also vanish.

The ordinary variation of the action gives rise to the following boundary
terms (where we have for the moment ignored the terms arising from the
ordinary variations of the fields in the vector multiplet as well
as the fields in the $\Sigma'$ and $P'$ multiplets)
\begin{eqnarray}
\lefteqn{
\delta_{\rm ord} (S_{\rm bulk} + S_{\rm bdry} ) = \int dx^0 
\left[ {i\over2} \left(\tau_i \delta \overline{\xi}_i + \overline{\tau}_i \delta
\xi_i \right) \right.
}
\nonumber \\
\lefteqn{-\left(D_1\overline{\phi}_i + 
|\varsigma|^2{{\partial E_a}\over{\partial \phi_i}} \overline{E}_a
- a_1\varsigma{{\partial^2
  E_a}\over{\partial\phi_i\partial\phi_j}}
\tau_j \overline{\pi}_a
+{{\partial E_a}\over{\partial \phi_i}}
\overline{\pi}_a \beta \right.} \nonumber \\
&&\left.\hspace{0.4in} +|p'|^2{{\partial J^a}\over{\partial \phi_i}} \overline{J}^a 
- a_2 p' {{\partial^2
  J^a}\over{\partial\phi_i\partial\phi_j}}
\tau_j \pi_a
+{{\partial J^a}\over{\partial\phi_i}} \pi_a \gamma
\right) \delta \phi_i 
\nonumber \\
\lefteqn{
-\left(D_1\phi_i + 
|\varsigma|^2{{\partial \overline{E}_a}\over{\partial \overline{\phi}_i}} E_a
- a_1\overline{\varsigma}{{\partial^2
  \overline{E}_a}\over{\partial\overline{\phi}_i\partial\overline{\phi}_j}}
\pi_a \overline{\tau}_j
+{{\partial \overline{E}_a}\over{\partial \overline{\phi}_i}}
\overline{\beta} \pi_a\right.}
\nonumber \\
&&\left.\hspace{0.4in}
+|p'|^2 {{\partial \overline{J}^a}\over{\partial \overline{\phi}_i}} J^a
- a_2 \overline{p}' {{\partial^2
  \overline{J}^a}\over{\partial\overline{\phi}_i\partial\overline{\phi}_j}}
\overline{\pi}_a \overline{\tau}_j
+{{\partial \overline{J}^a}\over{\partial \overline{\phi}_i}}
\overline{\gamma}\overline{\pi}_a
\right) \delta \overline{\phi}_i \nonumber  \\
&&+ \left( 
{i\over2}\overline{\xi}_i 
- a_1 \overline{\pi}_a 
\varsigma{{\partial E_a}\over{\partial\phi_i}}
- a_2 \pi_a p' {{\partial J^a}\over{\partial\phi_i}} \right) \delta \tau_i 
\nonumber \\
&&
+ \left({i\over2}\xi_i 
+ a_1 \pi_a \overline{\varsigma}
{{\partial \overline{E}_a}\over{\partial\overline{\phi}_i}}
+ a_2 \overline{\pi}_a \overline{p}'
{{\partial \overline{J}^a}\over{\partial\overline{\phi}_i}}
 \right) \delta \overline{\tau}_i \nonumber   \\
&& +\delta \overline{\pi}_a \left( i \widetilde{D}_0\pi_a -E_a\beta
+\overline{J}^a \overline{\gamma}
-a_1
\varsigma{{\partial E_a}\over{\partial\phi_i}} \tau_i +a_2
\overline{p}'
{{\partial \overline{J}^a}\over{\partial\overline{\phi}_i}} \overline{\tau}_i
\right) \\
&&\left.-\delta \pi_a \left( -i \widetilde{D}_0\overline{\pi}_a
-\overline{E}_a\overline{\beta} + J^a\gamma
+a_1 \overline{\varsigma}
{{\partial \overline{E}_a}\over{\partial\overline{\phi}_i}} \overline{\tau}_i
- a_2 p' {{\partial J^a}\over{\partial\phi_i}} \tau_i 
\right)\right]\nonumber 
\end{eqnarray}

Boundary conditions consistent with supersymmetry and the ordinary
variation of the total action are obtained when
$a_1=a_2={1\over2}$. We also add an additional boundary
contact term 
\begin{equation}
S_1^c = \int dx^0 i \left({{\sigma -
\overline{\sigma}}\over{\sqrt2}} \right) \left(\sum_i Q_i |\phi_i|^2
-r\right)
\end{equation}
The coefficient of $r$ is added for the
cancellation of the boundary terms in the vector multiplet sector which
we consider later.
In the matter sector, the boundary conditions that we obtain are
\begin{eqnarray}
&&\xi_i -i \overline{\varsigma}
{{\partial \overline{E}_a}\over{\partial\overline{\phi}_i}} \pi_a 
-i\overline{p}'{{\partial \overline{J}^a}\over{\partial\overline{\phi}_i}} 
\overline{\pi}_a =0 \label{genbca} \\
\lefteqn{
\widetilde{D}_1 \phi_i + {\partial\over{\partial\overline{\phi}_i}} 
\left( |\varsigma|^2|E_a|^2 + |p'|^2|J^a|^2\right) 
}\hspace{0.5in}
  \nonumber \\
&&- \overline{\varsigma}
{{\partial^2 \overline{E}_a}\over{\partial\overline{\phi}_i\partial\overline{\phi}_j}} 
\pi_a \overline{\tau}_j 
-\overline{p}'
{{\partial^2 \overline{J}^a}\over{\partial\overline{\phi}_i
\partial\overline{\phi}_j}} 
\overline{\pi}_a \overline{\tau}_j
+{{\partial \overline{E}_a}\over{\partial \overline{\phi}_i}}
\overline{\beta} \pi_a
+{{\partial \overline{J}^a}\over{\partial \overline{\phi}_i}}
\overline{\gamma}\overline{\pi}_a
=0  \label{genbcb}
\\
&&\overline{F}_i - i \varsigma p' \left.{\partial\over{\partial\phi_i}} \left({E}_a
{J}^a \right) \right.
=0 \quad,
\end{eqnarray}
where $\widetilde{D}_1 \phi_i \equiv [\partial_1 + iQ_i v_1 -iQ_i
{{\sigma-\overline{\sigma}}\over\sqrt2}]\phi_i$.
The last equation can be integrated to obtain the condition
\begin{equation}
i \varsigma p' E_a J^a = W
\end{equation}
when one has a superpotential. This is
different from the condition $E_a J^a =0$ seen in the
corresponding $(0,2)$ construction of vector bundles
for the heterotic string.  
We deal with this condition by introducing a chargeless spectator 
boundary fermi multiplet $\widehat{\Pi}$ satisfying
$$
\overline{\cal D} \widehat{\Pi} = 1
$$
with superpotential given by $\widehat{J}=W$.  Note that we {\it do not}
introduce any $\varsigma$ or $p'$ fields with this boundary fermion.
This is a solution proposed by Warner in the LG
context. Note that the $1$ above is actually a dimensionful scale $\mu$
which we have set to one. Thus the rest of the boundary fermions
will satisfy $E\cdot J=0$ as in the $(0,2)$ vector bundle
constructions for the heterotic string. Since this spectator multiplet
occurs for all examples, we will assume that this has been introduced
in all examples that we consider in this paper.

The equation of motion for the boundary fermions are
\begin{equation}
\widetilde{D}_0 \pi_a + i
\varsigma{{\partial E_a}\over{\partial\phi_i}} \tau_i
-i \overline{p}'{{\partial \overline{J}^a}\over{\partial\overline{\phi}_i}}
\overline{\tau}_i+iE_a \beta   
-i \overline{J}^a \overline{\gamma} =0
\end{equation}
One can now verify that all the boundary terms vanish in the low-energy limit
provided the boundary fermions are uncharged as we assumed earlier.

We now consider the  fields in the vector multiplet. 
On substituting for the $\xi_i$ using the boundary condition as derived earlier,
we obtain 
\begin{equation}
{{\sigma - \overline{\sigma}}\over{\sqrt2 i}} ={1\over{2K[\phi]}} \sum_i
Q_i\left[ 
\overline{\varsigma}
{{\partial \overline{E}_a}\over{\partial\overline{\phi}_i}}
\overline{\tau}_i \pi_a 
-\varsigma{{\partial E_a}\over{\partial\phi_i}} \tau_i \overline{\pi}_a
+ \overline{p}' {{\partial \overline{J}^a}\over{\partial\overline{\phi}_i}}
\overline{\tau}_i \overline{\pi}_a 
-p'{{\partial J^a}\over{\partial\phi_i}} \tau_i \pi_a
\right] 
\end{equation}
It is easy to see that this leads to the boundary condition
\begin{equation}
\sigma - \overline{\sigma}= 0
\end{equation}
on using the low-energy condition $\varsigma=p'=0$. The above equation
and its supersymmetric partners will be  the boundary
conditions on the fields in the vector multiplet in the GLSM.

The equation of motion for the fields $\varsigma$ and $\beta$ are
\begin{eqnarray}
\beta + \overline{E}_a \pi_a = 0 \\
i\tilde{D}_0 \varsigma +\varsigma |E_a|^2+ 
{{\partial\overline{E}_a}\over {\partial\overline{\phi}_i}}
\overline{\tau}_i \pi_a =0 
\end{eqnarray}
Similarly, the equations of motion  for $p'$ and $\gamma$ are
\begin{eqnarray}
i\tilde{D}_0 p' + p' |J^a|^2 +
{{\partial \overline{J}^a}\over {\partial\overline{\phi}_i}}
\overline{\tau}_i \overline{\pi}_a =0 \\
\gamma + \overline{J}^a \overline{\pi}_a = 0
\end{eqnarray}

\subsubsection{Examples}
\begin{enumerate}
\item A six-brane wrapping a Calabi-Yau threefold given by
a hypersurface $G=0$ in weighted projective space. This is given by
a single boundary fermi multiplet with $J=P' P$.  This fermion
has support on the space $p=0$. The restriction to the hypersurface (in
the NLSM) is implemented by continuity from the bulk.
\item A four-brane given by a holomorphic equation
$f(\phi)=0$ is described by an additional holomorphic constraint 
by  $J_1=f(\Phi)$ over and above the six-brane condition
discussed above. This $J_1$ comes with a $P_1'$ superfield.
Other lower dimensional branes can be obtained by the transverse
intersection of holomorphic conditions. This will introduce
as many $P'$-fields as there are conditions. 
The condition in
the vector multiplet remains unchanged from that of the
six-brane\footnote{This is a more satisfactory description of
lower-dimensional branes than the one chosen in \cite{lsmone}
where one chooses $f(\phi)=0$ as the boundary condition. In
this construction, this occurs as a low-energy condition.}.
This follows from the fact that $\pi$ vanishes when $f\neq0$ 
and as a consequence, $p'$ vanishes when $f\neq0$. It follows
that the combination $p'f$ vanishes in the NLSM limit.
\item  The restriction of $\Omega^1(1)$ to the quintic hypersurface.
where we have chosen the $J$'s as in section \ref{secvbpn}.
\item Consider the case of a vector bundle $(-V_2)$ (in then notation of
sec. \ref{secvbwpn}) on a degree six hypersurface
in $\BP^{1,1,1,1,2}$.  
In this case $J_i=\phi_i$ (in the notation of section  \ref{secvbwpn}). 
\end{enumerate}

\subsection{$\theta=0$ in the GLSM}

In order to lift the boundary conditions of the previous subsection
to the GLSM,
we have to deal with the boundary terms given by the ordinary
variations of fields in the bulk vector multiplet. We also
have to deal with the contributions to the boundary action of
${\cal O}({1\over{e^2}})$. 
The boundary terms involving fields in the vector multiplet
that arise from the ordinary variations of the 
bulk as well as $S_1^c$ are
\begin{eqnarray*}
\delta_{\rm ord} (S_{\rm bulk}+S_1^c) &=& {1\over{e^2}} \int dx^0 \left [
-(v_{01} + {{\theta e^2} \over{2\pi}})\ \delta v_0 \right. \\ 
&&- \partial_1\left({{\sigma + \overline{\sigma}}\over{\sqrt2}}
\right) \delta \left({{\sigma + \overline{\sigma}}\over{\sqrt2}} \right)
-i\widetilde{D} 
\delta \left({{\sigma - \overline{\sigma}}\over{\sqrt2}} \right)   \\
&& +{i\over2}\left(
\left({{\overline{\lambda}_- -
\overline{\lambda}_+}\over{\sqrt2}}\right)
\delta \left({{\lambda_- + \lambda_+}\over{\sqrt2}}\right)
+ \left({{\overline{\lambda}_+ +
\overline{\lambda}_-}\over{\sqrt2}}\right)
\delta \left({{\lambda_- - \lambda_+}\over{\sqrt2}}\right)\right. \\
&&\left.\left.+ \left({{\lambda_- - \lambda_+}\over{\sqrt2}}\right)
\delta \left({{\overline{\lambda}_+ + 
\overline{\lambda}_-}\over{\sqrt2}}\right)
+ \left({{\lambda_+ + \lambda_-}\over{\sqrt2}}\right)
\delta \left({{\overline{\lambda}_- - 
\overline{\lambda}_+}\over{\sqrt2}}\right)
\right)  \right]
\end{eqnarray*}
There is no contribution to the above from $S^1_{\rm bdry}$ since we have
chosen $\pi_a$ to be chargeless. However, there is a contribution
to $\delta \tilde{v}_0$ coming from $S^2_{\rm bdry}$ i.e., the kinetic part of
the boundary action involving the $\Sigma'$ and $P'$ superfields.

First, we choose
\begin{eqnarray}
\sigma -\overline{\sigma}= 0 \\
\lambda_+ - \eta \overline{\lambda}_- = 0 \\
v_{01}- \eta \partial_1\left({{\sigma +
\overline{\sigma}}\over{\sqrt2}}\right)=0
\end{eqnarray}
With the above choice of boundary conditions, most of boundary terms in
the ordinary variations of $S_{\rm bulk} + S_1^c$ vanish except for the
terms involving $\delta \tilde{v}_0$. The coefficient of this term is
$$
-{v_{01}\over{e^2}} + Q_\varsigma |\varsigma|^2 + Q_{p'} |p'|^2\quad.
$$
Thus, the final boundary condition is
\begin{equation}
{v_{01}\over{e^2}} = Q_\varsigma |\varsigma|^2 + Q_{p'} |p'|^2\quad.
\end{equation}
These boundary conditions when combined with the  boundary conditions 
on the matter fields and equations  of motion for the boundary fields
(as obtained earlier in the low-energy limit)
lead to a complete cancellation of all boundary terms which occur in
both the ordinary as well as supersymmetric variation of the action.
The case of multiple $P'$ fields easily generalises and we shall not
discuss them any more.

It is useful to comment here that 
the equations of motion for all the boundary fields 
($\pi_a$, $p'$ and $\sigma$) can all be obtained from superfield
actions. This  reflects the fact that all boundary terms which
appear in the supersymmetric variation of the complete
bulk and boundary actions vanish.  For instance, for the $\pi$ field
we had chosen $a_1=a_2=1/2$ when the superfield expansion
gives $a_1=a_2=1$ in the action. The equation of motion for $\pi$
however is as if we had chosen the latter.

\subsection{An intriguing observation}

So far, we have implicitly restricted our attention to  the one-particle
sector of the boundary fermion Fock space as we indicated in the
introduction. This restriction was essential to obtain the vector bundle
of interest rather than all its antisymmetric powers.   One may wonder
whether sectors of other fermion number have any meaning. In this
regard, consider the case of the  cotangent bundle to $\BP^4$ that we
discussed earlier. The boundary state associated with it can be obtained
as
\begin{equation}
|B\rangle \sim \int [D\pi]\ e^{ iS_{\rm bdry}}\  {\cal P}_1\  |B\rangle_0
\otimes |0\rangle_\pi
\end{equation}
where we have schematically indicated the realisation of the boundary
state associated with the cotangent bundle as the path-integral
over all the boundary fields (symbolically indicated by $[D\pi]$)
subject to the restriction to one-particle
states (shown by the projector ${\cal P}_1$). By $|B\rangle_0$, we mean
the state associated with Neumann boundary conditions on all
matter fields. 

If instead, in the boundary state, we  change the
restriction to fermion number $p$, the index computation suggests that
on obtains the  boundary state associated to the vector bundle
$\Omega^p(p)$. These states form the large-volume analogue of
the Recknagel-Schomerus states in the $\sum_a l_a=0$. Pushing this
further, to more complicated examples involving weighted projective
spaces such as $\BP^{1,1,1,1,2}$, we expect to  recover the $\sum_a l_a=0$
orbit in such cases provided one suitably modifies
the projection on states with definite particle number.

However, under large-volume monodromy as implemented in this paper, it
is not clear that all $\Omega^p(p)$ will behave in an appropriate
fashion.  Hence, we prefer to realise $\Omega^p(p)$ only via the
one-particle projection involving  a GLSM realisation of
complexes of suitable length.

\section{Charged fermions, $\theta\neq0$ and large volume monodromy}
\label{seclvmon}

As has been shown in a simple example in\cite{lsmone,HIV}, the inclusion of the
$\theta$ term
requires the addition of contact terms (derivable in the NLSM).
This modified the boundary conditions in a manner which was compatible
with supersymmetry. For
the case of vector bundles on Calabi-Yau manifolds, we will need an
an additional condition: Under $\theta\rightarrow \theta+2\pi$, the
monodromy of the B-branes around the singularity at large volume
should be implemented correctly. It turns out
that the monodromy corresponds to tensoring the bundle with a
line bundle (${\cal O}(-1)$ for the quintic). This process does not
affect the stability and the moduli space of the bundle at large volume.
As we shall see, this has a simple and elegant realisation in the
NLSM limit. 

In the monad construction, the large-volume monodromy action
corresponds to shifting the charges of all boundary fermions by one
unit. Thus, we can anticipate that dealing with the case of charged
boundary fermions should be closely related to turning on the theta
term.
When the fermions are charged, the 
the boundary terms in the ordinary variation do not vanish 
in the NLSM limit. This is
due to the term of the form $Q_\pi\overline{\pi}\pi \delta \widetilde{v}_0$.
This is very similar to the boundary term which appears when
one turns on the theta term i.e., $\theta\delta v_0/2\pi$. Hence,
we expect a contact term involving bilinears of the boundary fermions
playing the role of $\theta$. Such contact terms, among other things,
modify the equations of motion of the boundary fermions.

\subsection{$\theta\neq0$ and the low-energy limit}

{}From large-volume monodromy considerations, when $\theta=-2\pi n$,
we expect the fermion to have charge $Q_\pi=n$ and the vector 
bundle $E$ becomes $E(n)=E\otimes {\cal O}(n)$. This corresponds to
turning on a $U(1)$
gauge field on the worldvolume of the brane. The 
coupling of the boundary fermions to the gauge field
takes the form (similar terms appear in the NLSM 
considerations in the context of the heterotic string\cite{callanreview})
\begin{equation}
S_{\rm gauge~field}= -\int dx^0\left\{ {{in}\over{2 r}} \overline{\pi}_a
\pi_a \sum_i (\phi_i \partial_0\overline{\phi}_i - \overline{\phi}_i
\partial_0 \phi_i) \right\}
\end{equation}
One can, in fact,  verify that this gauge field shifts the first Chern
class of the bundle in an appropriate fashion (see \cite{HIV} for
the case when $E$ is a line bundle). The boundary fermions 
also couple to the bulk fermions $\tau_i$ and $\overline{\tau}_i$.

Let us add the above term to the boundary action. The modification to
boundary condition on the bulk scalar fields is clear from above. 
One obtains the following boundary condition
from the cancellation of boundary terms proportional
to $\delta\overline{\phi}$ in the ordinary variation
\begin{equation}
\widetilde{D}_1 \phi_i +
{{in}\over{ r}}\overline{\pi}_a \pi_a  \partial_0\phi_i 
+{{in}\over{2 r}}\partial_0 (\overline{\pi}_a \pi_a)  \phi_i 
+ \cdots =0 \label{formone}
\end{equation}
where the ellipsis denotes terms which are theta independent.
Such a term can arise from the supersymmetric variation of
the following boundary condition on the matter fermions of
the form
\begin{equation}
\xi_i - {{in}\over{2 r}} \overline{\pi}_a\pi_a  \tau_i  + \cdots =0 
\end{equation}
whose supersymmetric variation leads to
\begin{equation}
\widetilde{D}_1 \phi_i +{{in}\over{ r}}\overline{\pi}_a
\pi_a  \widetilde{D}_0\phi_i 
- {{in}\over{2 r}} \delta_{susy}(\overline{\pi}_a\pi_a) \tau_i
+ \cdots =0 \label{formtwo}
\end{equation}
where by $\delta_{susy}(\overline{\pi}_a\pi_a)$, we mean the
term proportional to $\overline{\epsilon}$ in the supersymmetric
variation of $\overline{\pi}_a\pi_a$. In order for the two
equations (\ref{formone}) and (\ref{formtwo}) to match\footnote{In the
NLSM limit, terms which lead to the $\widetilde{v}_0$ term in
(\ref{formtwo}) vanish in the ordinary variations on using
$\delta(\sum_i \overline{\phi}_i\phi_i=0)$.}), we need
$$
\delta_{susy}(\overline{\pi}_a\pi_a)=0\quad{\rm and}\quad
\partial_0(\overline{\pi}_a\pi_a)=0
$$
It turns out that both expectations are true in the low-energy/NLSM limit
on using the equations of motion of $\pi_a$ which turn out
to be of the form
$$
i\partial_0 \pi_a -{{in}\over{ r}}\sum_i (\phi_i
\partial_0\overline{\phi}_i - \overline{\phi}_i
\partial_0 \phi_i) + {n\over{r}} \sum_i\overline{\tau}_i\tau_i + \cdots
=0
$$
The above equation clearly reflects the fact that we have indeed turned
on a gauge field on the worldvolume of the brane. In the NLSM limit,
this is in fact equivalent to  the following equation
$$
i\widetilde{D}_0 \pi_a + \cdots =0
$$
where we have introduced a covariant derivative corresponding to
a fermion of charge $n$. This shows that {\em gauge invariance} as well
as {\em supersymmetric invariance} necessarily  forces the change of
boundary fermion charge and matches the monad construction for
the vector bundle $E(n)$.

Keeping in mind that we are interested in the GLSM, we rewrite
the boundary term corresponding to turning on a gauge field
in the NLSM as
\begin{equation}
S_{\rm gauge~field}= \int dx^0\left\{ 
-{{in}\over{2 r}} \overline{\pi}_a \pi_a
\sum_i (\phi_i \widetilde{D}_0\overline{\phi}_i - \overline{\phi}_i
\widetilde{D}_0 \phi_i)
\right\}
\end{equation}
with an additional condition that the charge of the boundary fermion
is shifted by $n$ by appropriate covariant derivatives in the boundary
action.

The boundary term which we add in the NLSM for non-zero theta takes the
form
\begin{equation}
S^{NLSM}_{\rm boundary}= \int dx^0\left\{  i{\Theta\over{2\pi r}}
\sum_i (\phi_i \widetilde{D}_0\overline{\phi}_i - \overline{\phi}_i
\widetilde{D}_0 \phi_i)
\right\}
\end{equation}
where 
$$
{\Theta\over{2\pi r}}\equiv \left[{{\theta_f}\over{2\pi r}} 
+{{[\theta/2\pi]}\over{2 r}} \overline{\pi}_a \pi_a\right]\quad.
$$
Here $[\theta/2\pi]$ is the integer part of $\theta/2\pi$ and
$\theta_f/2\pi$ is the fractional part of $\theta/2\pi$.

The boundary conditions (for non-zero $\theta$) is modified to
\begin{eqnarray}
&&\xi_i - i {\Theta\over{2\pi r}} \tau_i -i \overline{\varsigma}
{{\partial \overline{E}_a}\over{\partial\overline{\phi}_i}} \pi_a 
-i\overline{p}'{{\partial \overline{J}^a}\over{\partial\overline{\phi}_i}} 
\overline{\pi}_a =0  \\
\lefteqn{
\widetilde{D}_1 \phi_i +i {\Theta\over{2\pi r}} \widetilde{D}_0\phi_i
+ {\partial\over{\partial\overline{\phi}_i}} 
\left( |\varsigma|^2|E_a|^2 + |p'|^2|J^a|^2\right) 
}\hspace{0.5in}
  \nonumber \\
&&- \overline{\varsigma}
{{\partial^2 \overline{E}_a}\over{\partial\overline{\phi}_i\partial\overline{\phi}_j}} 
\pi_a \overline{\tau}_j 
-\overline{p}'
{{\partial^2 \overline{J}^a}\over{\partial\overline{\phi}_i
\partial\overline{\phi}_j}} 
\overline{\pi}_a \overline{\tau}_j
+{{\partial \overline{E}_a}\over{\partial \overline{\phi}_i}}
\overline{\beta} \pi_a
+{{\partial \overline{J}^a}\over{\partial \overline{\phi}_i}}
\overline{\gamma}\overline{\pi}_a
=0 
\\
&&\overline{F}_i - i \varsigma p' \left.{\partial\over{\partial\phi_i}} \left({E}_a
{J}^a \right) \right.
=0 \quad,
\end{eqnarray}
The equation of motion for the boundary fermions get modified to
\begin{equation}
\widehat{D}_0 \pi_a 
+ i \varsigma{{\partial E_a}\over{\partial\phi_i}} \tau_i
-i \overline{p}'{{\partial \overline{J}^a}\over{\partial\overline{\phi}_i}}
\overline{\tau}_i+iE_a \beta   
-i \overline{J}^a \overline{\gamma} =0
\end{equation} \label{pieqn}
where $\widehat{D}_0 \pi_a = ( \partial_0 + i [{\theta\over{2\pi}}]
\widetilde{v}_0)\pi_a$ 
reflects the  shift in the charge of $\pi_a$ from $0$ to
$[{\theta\over{2\pi}}]$. Further, from eqn. (\ref{stgaugefield}) we can see
that $\widetilde{v}_0$ is the pullback of a spacetime  $U(1)$ gauge field of
constant field strength to the worldvolume of the  D-brane.

One can now verify, that  for the above choice of boundary conditions
and eqn. of motion for the boundary fermion, all boundary terms in the
ordinary as well as supersymmetric variation vanish (in the NLSM limit).
We will now obtain boundary conditions on the fields in the bulk vector
multiplet. The boundary conditions on the the fields $\sigma$ and
$\overline{\sigma}$ are
\begin{equation}
{{\sigma - e^{-2i\gamma} \overline{\sigma}}\over{\sqrt2 i}} 
=0
\end{equation}
where $\tan \gamma =-{\Theta\over{2\pi r}}$.

\subsection{$\theta\neq0$ in the GLSM}

In the NLSM limit, the consistency of the boundary conditions with both
the ordinary and supersymmetric variation required
$\delta_{susy}(\overline{\pi}_a\pi_a)=0$ and
$\partial_0(\overline{\pi}_a\pi_a)=0$ which was true in the NLSM on using
the equations of motion for $\pi$. We
will show that something similar occurs in the GLSM. We will
require a ${\cal J}$ which has similar properties i.e.,
$$
\delta_{susy}({\cal J})=0\quad;\quad \partial_0 ({\cal J})=0
\label{curlyJ} \quad.
$$
It is easy to see using the equations of motion (of the boundary fields)
that ${\cal J}$ is  given by\footnote{We note here that our 
choice of a first-order action for $\varsigma$ and $p'$ fields leads to 
a linear coupling to $\widetilde{v}_0$ for all boundary fields. This plays
an important role in ensuring a simple and closed form of ${\cal J}$ with
the required properties.}
\begin{equation}
{\cal J} \equiv \left (\overline{\pi}_a \pi_a - |\varsigma|^2 + |p'|^2
\right)
\end{equation}
This reflects the fact that we expect that the only change in
the boundary equations  of motion is through the change in the 
charge of the fields. Such changes obviously do not effect $\partial_0
{\cal J} =0$ since it has zero charge.

We define 
$$
{\Theta\over{2\pi r}}\equiv \left[{{\theta_f}\over{2\pi r}} 
+{{[\theta/2\pi]}\over{2 r}} {\cal J}\right]\quad.
$$
and choose the theta dependent boundary (contact) term to be
\begin{equation}
S^{GLSM}_{\rm boundary}= \int dx^0\left\{  i{\Theta\over{2\pi r}}
\sum_i (\phi_i \widetilde{D}_0\overline{\phi}_i - \overline{\phi}_i
\widetilde{D}_0 \phi_i)
\right\}
\end{equation}
in addition to the {\em appropriate shift in charge} for the boundary fields
$\pi$, $p'$ and $\varsigma$.

It is straightforward to obtain the equations of motion for the
boundary fields. For example, the equation of motion for $\pi_a$
is
\begin{equation}
\widehat{D}_0 \pi_a + i
\varsigma{{\partial E_a}\over{\partial\phi_i}} \tau_i
-i \overline{p}'{{\partial
\overline{J}^a}\over{\partial\overline{\phi}_i}}
\overline{\tau}_i+iE_a \beta
-i \overline{J}^a \overline{\gamma} +{\cal O}(1/e^2)=0
\end{equation}
where 
$$
\widehat{D}_0 \pi_a = \partial_0 \pi_a + i [\theta/2\pi]
\left(\widetilde{v}_0 + {i\over 2}\sum_i Q_i(\phi_i
\widetilde{D}_0\overline{\phi}_i - \overline{\phi}_i
\widetilde{D}_0 \phi_i) - \overline{\tau}_i\tau_i \right) \pi_a \quad.
$$
There are three new contributions which arise in comparison with the
$\theta=0$ case. The $\widetilde{v}_0$ is the piece arising from
the change in charge, the $\overline{\tau}_i\tau_i$ piece comes from
from the ordinary variations of $\delta\xi$ and $\delta\overline{\xi}$.
The third piece comes from the boundary term that we add. It is easy to
see that in the NLSM limit, on using the D-term constraint, that the
three pieces collapse precisely to the pull-back of a constant gauge
field as in the line-bundle case\cite{HIV}. The ${\cal O}(1/e^2)$
indicates potential contributions from the vector multiplet sector  which we
have not included.

The boundary conditions on the bulk fields in the vector multiplet
now follow straightforwardly
from the calculations of \cite{lsmone}, with the sole modification that
the $\theta$ of the earlier paper is replaced appropriately by the
$\Theta$ that we have defined above. We write down the relevant
equations below:

\begin{eqnarray}
({\bar\sigma}-e^{2i\gamma}\sigma)|_{x^1=0}&=&0\\
({\lambda_{+}}-\eta e^{2i\gamma}\lambda_{-})|_{x^1=0}&=&0 \\
\left.\left(v_{01}+
{\Theta\over{2\pi r}}D\right)\right|_{x^1=0}&=&Q_\varsigma |\varsigma|^2
+ Q_{p'} |p'|^2 + Q_\pi \overline{\pi}_a \pi_a \\
\left.\left(\partial_1{{\overline{\sigma} + e^{2i\gamma}
\sigma}\over{\sqrt2}}
-\eta {{\Theta}\over{2\pi r}} 
e^{i\gamma} D\right)\right|_{x^1=0} &=&Q_\varsigma |\varsigma|^2 +
Q_{p'} |p'| + Q_\pi \overline{\pi}_a \pi_a 
\end{eqnarray}
where $Q_\pi=[\theta/2\pi]$, $Q_\varsigma = Q_\pi - {\rm deg}(E)$ and
$Q_{p'}=-Q_\pi -{\rm deg}(J)$.

\section{Vector bundles from complexes of length $>2$}
\label{seclong}

In the field theoretic construction, it seems most natural to consider
complexes whose elements are direct sums of line bundles rather than
vector bundles. This is however not a very restrictive condition.
For example, all coherent sheaves on $\BP^n$ (and those on
weighted projective spaces)
can be obtained as the cohomology of such complexes. This is a result due
to Beilinson.

Before getting to the most general situation, it is useful to study an
example which naturally illustrates how one deals with longer complexes.
The simplest one is that of $\Omega^2(2)$ which we shall discuss now.
\subsection{$\Omega^2(2)$ and $\BT^2(-2)$}
As was discussed earlier, the following exact sequence gives rise
to $\Omega^2(2)$.
\begin{equation}
0\rightarrow \Omega^{2}(2)
\rightarrow {\cal O}^{\oplus 10}\stackrel{J_{[ij]}^k}{\longrightarrow}
\Omega^{1}(2)\rightarrow 0
\end{equation}
Following the discussion in the earliers sections,
we consider ten fermi multiplets $\Pi^{[ij]}$ with the
superpotential
\begin{equation}
S_J=-{1\over\sqrt2} \int dx^0 d\theta \left(\Pi^{[ij]}
J_{[ij]}^k(\Phi')P_k'\right)|_{\overline{\theta}=0} -
{\rm h.c.}
\end{equation}
where $J_{[ij]}^k(\phi) = (\phi_i \delta_j^k - \phi_j \delta_i^k)$ and
$P_k'$ are five boundary chiral multiplets. 

It is easy to see that
the above superpotential admits the gauge invariance (with bosonic
gauge parameter $b$)
\begin{equation}
p'_k \sim p'_k + b \phi_k 
\end{equation}
which is implied by the identity:  $\phi_k J_{[ij]}^k(\phi)=0$.  
Thus, even though the superpotential gives mass to four fermions
in the fermi multiplet, the gauge invariance indicates that
one linear combination of the $P'_k$ remain massless. 
One has to fix this gauge invariance which can be done by
imposing the following constraint of the superfield $ P'_k$
\begin{equation}
\overline{\cal D} P'_k = \sqrt2  N \Phi'_k
\end{equation}
where $N$ is a chiral fermi superfield(with lowest component $n$). 
It is easy to see that $P'_k$ is now
a section of $\BT^1(-2)=(\Omega^1(2))^*$ given by tensoring the Euler
sequence with ${\cal O}(-1)$:
\begin{equation}
0\rightarrow {\cal O}(-2)
\rightarrow {\cal O}^{\oplus 5}(-1)\rightarrow
\BT^1(-2)\rightarrow 0
\end{equation}
Thus, in the GLSM construction, one is  implementing
the following exact sequence
\begin{equation}
0\rightarrow \Omega^{2}(2)
\rightarrow {\cal O}^{\oplus 10}\rightarrow
{\cal O}^{\oplus 5}(1)\rightarrow
{\cal O}(2)\rightarrow 0 
\end{equation}
One can now verify that $P'_k$ being a section of $\BT^1(-2)$ is
consistent with the superpotential $S_J$ being a scalar. 
This also explains
how a holomorphic constraint (in the sequence given above) appears
as a gauge invariance in the GLSM construction.

An important point to note that we need to choose a {\em first order
action} for the bosonic field $p'_k$. This is essential for obtaining the right
number of massless fermions in the NLSM limit. Thus, the kinetic energy
that we choose for the superfield  $P'_k$ is
\begin{equation}
S_{P'_k} = \int dx^0 d^2\theta \overline{P}'_k P'_k
\end{equation}
which leads to the mass term $\gamma_k \overline{n} \overline{\phi}_k$ for the
remaining massless fermion in $P'_k$. If we had chosen a standard
second order action for the bosons, one can verify that the expected
mass term does not appear. Thus, one is forced to choose the first
order action. We also include a standard action for the $N$ superfield. 

The $p'_k$ fields thus behave like ``ghost fields'' that appear whenever
one has nested gauge-invariances. Their action is like that of a fermion
but their statistics are bosonic. The determinant associated with
the partition function of one such bosonic ghost is cancelled by that
a fermion. 
It is useful to ``count'' the boundary chiral superfields that we have 
introduced:
there are eleven fermi superfields and five bosonic chiral superfields
which leads effectively to six (massless)
fermi superfields (this is equal to the dimension of $\Omega^2(2)$). 

A related example is the construction for $\BT^2(-2)$. This is obtained
by considering ten fermi superfields $\Pi_{[ij]}$ subject to the
constraint 
$$ 
\overline{\cal D} \Pi_{[ij]} = \sqrt2  \Sigma_k
E_{[ij]}^k(\Phi')\quad.
$$
where $\Sigma_k$ are five superfields subject to the constraint
$$
\overline{\cal D} \Sigma_k = \sqrt2 N E_k(\Phi') \quad.
$$
Here we have introduced a chiral fermi superfield $N$. The consistency
condition between the two constraints is
$$
\sum_k E_{[ij]}^k(\Phi')\  E_k(\Phi') =0
$$
which is statement that the composition of two consecutive maps in
the following complex vanish.
\begin{equation}
0\rightarrow {\cal O}(-2) \stackrel{E_k}{\rightarrow} {\cal O}(-1)^{\oplus5} 
\stackrel{E_{[ij]}^k}{\longrightarrow}
{\cal O}^{\oplus 10} \rightarrow \BT^2(-2)\rightarrow0
\end{equation}

\subsection{The general case}

In the general situation, the complex may have longer length and also
the cohomology may occur in more than one place. First, the generalisation
to complexes of arbitrary length is fairly straightforward. One begins by
introducing fermi multiplets at the point where the cohomology occurs. At
other points one introduces bose or fermi multiplets depending on the 
position in the complex. The charges of the fields are fixed by the line
bundles which occur at the each point. Finally, one has to fix whether
a given map is implemented through a superpotential (holomorphic
constraint) or through the gauge fixing of a gauge invariance. 

The situation, where  the cohomology occurs at two places in the complex
(separated by even number of terms) also goes through in a similar fashion.
The massless fermions that appear will arise from fields introduced at
these two places instead of one as in the examples that we considered.

It clearly of interest to extend this construction to
the case of vector bundles in the heterotic string. The major difference
is the change in dimension -- here we were dealing with a quantum
mechanical situation while in the heterotic string
we have a $1+1$ dimensional case. Thus, naively our first order actions
do not work in this case. But naive arguments do not constitute a no-go
theorem and the issue remains open.

\section{Bound states}

\subsection{D4-brane}

Let us recall the the construction for a (complex) codimension one brane
i.e., the D4-brane for the quintic. 
Let us consider the case when the D4-brane is given
by $\phi_1=0$. In the construction of this paper, one considers the
following sequence
\begin{equation}
0\rightarrow {\cal O} \stackrel{\phi_1}{\rightarrow}{\cal
O}(1)\rightarrow {\cal O}_H \rightarrow 0
\end{equation}
where ${\cal O}_H$ is the sheaf with support on the hyperplane
$\phi_1=0$.  Since ${\rm ch}({\cal O}_H) = {\rm ch} {\cal O}(1)- {\rm ch}
{\cal O}$, using the fact that Chern character preserves K-theory classes,
one can reinterpret this construction to be the {\em bound state}
of the D-brane corresponding to ${\cal O}(1)$ and the anti-brane
corresponding to ${\cal O}$.  This observation (to our knowledge) first
appeared in \cite{bpsalgebra}. For more recent work, see 
\cite{Kutasov:2000aq,Kraus:2000nj,hori} and references therein
In our construction, the boundary Lagrangian for this is
given by (excluding kinetic energy pieces for $\varsigma$ and $\beta$)
\begin{equation}
S_{\rm bdry} = \int dx^0 \left(
i\overline{\pi}\widetilde{D}_0 \pi - 
|\varsigma|^2|\phi_1|^2 
- \phi_1\overline{\pi}_a \beta
- \overline{\phi}_1 \overline{\beta} \pi
- a_1 \overline{\pi}
\varsigma\tau_1
- a_1 \overline{\varsigma}
\overline{\tau}_1 \pi \right)
\end{equation}
The terms that appear in this Lagrangian are reminiscent of the
Atiyah-Bott-Shapiro construction as discussed in 
\cite{Kutasov:2000aq,Kraus:2000nj,hori}. Following the discussion
in \cite{quintic}, we can see that a general hyperplane condition
of the form $\sum_i a_i\phi_i=0$ has four parameters. These can
be considered to be four possible deformations of the $\phi_1=0$
condition for the D4-brane.

More generally, suppose a brane associated with
a vector bundle $E_3$ is expected to arise as the
bound state of a brane (associated with vector bundle $E_1$)
 and an anti-brane (of the brane associated with bundle $E_2$).
(We assume that all three vector bundles have rank greater than
zero.)
First, conservation of RR charge implies that ${\rm ch} E_3
= {\rm ch} E_1 - {\rm ch} E_2$. This can occur in one of the following
two sequences
\begin{eqnarray*}
0 \rightarrow E_3 \rightarrow E_1 \rightarrow E_2 \rightarrow 0 \\
0 \rightarrow E_2 \rightarrow E_1 \rightarrow E_3 \rightarrow 0 
\end{eqnarray*}
This ambiguity is resolved by using the condition that the bundle $E_1$
is semi-stable\cite{bpsalgebra}.

\subsection{$\sum_a l_a =1$ as a bound state}
We have so far considered the large-volume analogues of $\sum_a l_a=0$
Recknagel-Schomerus states\cite{quintic}. We will now consider the
$\sum_a l_a=1$ states. At the Gepner point, these states are bound states
of two $\sum_a l_a=0$ RS states. Let us consider the vector bundle
${\cal B}$ given by the complex
\begin{equation}
0\rightarrow {\cal B} \rightarrow \Omega^1(1) \rightarrow {\cal O}
\rightarrow 0
\end{equation}
The D-brane corresponding to ${\cal B}$ has four moduli. Our strategy will
be to first construct the bound state in $\BP^4$ and then restrict to the
Calabi-Yau manifold. We further assume that no new moduli appear on
restriction (this can be actually proven).

In the GLSM, we constructed $\Omega^1(1)$ by considering five fermi 
multiplets $\Pi^i$ subject to the degree one holomorphic constraint $\phi_i
\pi^i=0$. In order to obtain the bound state ${\cal B}$, we must
impose an additional degree zero constraint:
\begin{equation}
a_i \pi^i =0\quad,
\end{equation}
where $a_i$ are five constants.  Given the overall scaling of the above
relation this leads to four independent parameters which we identify
with the moduli. We will show that ${\cal B}$ is a sheaf on $\BP^4$.
Consider the case when $a_1\neq0$. We fix the scaling by setting
$a_1=1$. In the chart $\phi_1=1$, it is easy to see that at the point
$(1,a_2,a_3,a_4,a_5)$, the two conditions collapse to one. However,
this point does not lie on the (generic) quintic hypersurface and hence
on restriction to the quintic hypersurface, we obtain a vector
bundle\footnote{We thank M. Douglas for a useful discussion on this
point}. The extension of this to other cases involves longer sequences
and can be worked out using the methods discussed in this paper.

\section{Summary}

We briefly recapitulate here the main results of the paper. 
First we
have given an explicit GLSM description of B-type D-brane configurations
that mathematically correspond to the monad construction of vecctor 
bundles. Secondly, we have shown how the correct large-volume 
monodromy action on these configurations requires the modification of 
the contact term at the boundary. Third, we extend the techniques of
(0,2) type constructions in the GLSM to the case where we have complexes
of length greater than two. Thus we have a complete description of all
D-brane configurations that are the large-volume analogues of the
$\sum_a l_a=0$ Recknagel-Schomerus states. Finally we show how D-branes
configurations that are bound
states can be naturally described in the GLSM and we show how the
appropriate moduli can be described. 

\noindent {\bf Acknowledgments} We thank L. Alvarez-Gaum\'e, M. Douglas,
K. Hori, K. Paranjape and T. Sarkar for useful discussions and
correspondence.

\end{document}